\documentclass[10pt,aps,prl,showpacs,floatfix,floats,superscriptaddress,twocolumn]{revtex4-1}
\usepackage{graphicx,latexsym}
\usepackage{dcolumn}
\usepackage{amssymb,amsmath,bm}

\usepackage[dvips]{color}

\usepackage{ulem}

\begin{document}

%-----------------------------------------------------------------

\title{Coupled collective and Rabi oscillations triggered by\\ electron transport through a photon cavity}

\author{Vidar Gudmundsson}
\email{vidar@hi.is}
\affiliation{Science Institute, University of Iceland, Dunhaga 3, IS-107 Reykjavik, Iceland}
\author{Anna Sitek}
\affiliation{Science Institute, University of Iceland, Dunhaga 3, IS-107 Reykjavik, Iceland}
\affiliation{Department of Theoretical Physics, Wroc{\l}aw University of Technology, 50-370 Wroc{\l}aw, Poland}
\author{Pei-yi Lin}
\affiliation{Science Institute, University of Iceland, Dunhaga 3, IS-107 Reykjavik, Iceland}
\author{Nzar Rauf Abdullah}
\affiliation{Science Institute, University of Iceland, Dunhaga 3, IS-107 Reykjavik, Iceland}
\author{Chi-Shung Tang}
\email{cstang@nuu.edu.tw}
\affiliation{Department of Mechanical Engineering, National United University, Miaoli 36003, Taiwan}
\author{Andrei Manolescu}
\email{manoles@ru.is}
\affiliation{School of Science and Engineering, Reykjavik University, Menntavegur 
             1, IS-101 Reykjavik, Iceland}

%
%----------------------------------------------------------------

\begin{abstract}
We show how the switching-on of an electron transport through a system
of two parallel quantum dots embedded in a short quantum wire in a
photon cavity can trigger coupled Rabi and collective electron-photon
oscillations. We select the initial state of the system to be an
eigenstate of the closed system containing two Coulomb interacting
electrons with possibly few photons of a single cavity mode.  The
many-level quantum dots are described by a continuous potential. The
Coulomb interaction and the para- and dia-magnetic electron-photon
interactions are treated by exact diagonalization in a truncated
Fock-space. To identify the collective modes the results are compared
for an open and a closed system with respect to the coupling to external
electron reservoirs, or leads. 
We demonstrate that the vacuum Rabi oscillations can
be seen in transport quantities as the current in and out of the system.
\end{abstract}


\maketitle
\textit{Introduction}.--Fine-tuning of the electron-photon interaction
has opened up new possibilities in semiconductor physics.
The transport of electrons through quantum dots assisted by up
to four photons in the teraherz frequency range has been observed
\cite{PhysRevLett.109.077401}, and double quantum dots have been
used to detect single-photons from shot-noise in electron transport
through a quantum point contact \cite{PhysRevLett.99.206804}. The
properties and control of atomic or electronic systems in photonic
cavities is a common theme in the research effort of many teams working
on various aspects of quantum cavity electrodynamics and related fields
\cite{PhysRevLett.50.1903,Gallardo:10,Frey11:01,Peterson380:2012,PhysRevB.86.085316,
PhysRevLett.110.013601,PhysRevLett.111.176401,PhysRevB.88.085429}.
The non-local single-photon transport properties
of two sets of double quantum dots within a photon cavity has recently been
modeled \cite{PhysRevB.87.195427}, and also a pump-probe
scheme for electron-photon dynamics in a hybrid conductor-cavity
system with one electron reservoir \cite{PhysRevB.90.085416}.
Many tasks in quantum information processing might be served by mixed
photon-electronics circuits.  In order to model such systems we need to
combine methods and tools that have traditionally been used and developed
in the fields of time-dependent electron transport and quantum optics.
In this publication we show how time-dependent electron transport
through a nanoscale system embedded in a photon cavity could be used
to detect vacuum Rabi-oscillations in it. In order to do so we use a
generalized master equation (GME) formalism for time-dependent electron
transport, that was initially developed for quantum optics systems
\cite{Zwanzig60:1338,Nakajima58:948}.

\begin{figure}[htbq]
      \includegraphics[width=0.38\textwidth,angle=0,bb=23 12 239 161,clip]{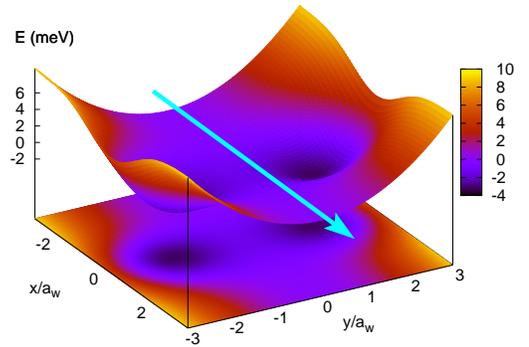}
      \caption{(Color online) The potential landscape defining the two parallel
               quantum dots in a short parabolically confined quantum wire.
               The arrow (cyan) indicates the general direction ($x$-direction) of electron transport
               after the system has been opened up.  
               The effective magnetic length $a_w=23.8$ nm, $\hbar\Omega_0=2.0$ meV, and $B=0.1$ T.}
      \label{W_2QD}
\end{figure}

\textit{The closed system in equilibrium}.--We consider a two-dimensional electron
system lying in the $xy$-plane (GaAs-parameters, $\kappa =12.4$,
and $m^*=0.067m_e$), subject to a homogeneous external weak magnetic
field in the $z$-direction ($B=0.1$ T). The system represents a short
quantum wire with parabolic confinement in the $y$-direction,
with energy $\hbar\Omega_0=2.0$ meV, but hard walls in the
$x$-direction. Two shallow parallel quantum dots are embedded in the
wire as is illustrated in Fig.\ \ref{W_2QD}. The external magnetic
field and the parabolic confinement define the natural length scale
$a_w=\sqrt{\hbar /(m^*\Omega_w)}$, with
$\Omega_w=\sqrt{\Omega_0^2+\omega_c^2}$, where $\omega_c=(eB/m^*c)$.
The Coulomb interaction of the electrons in the system is considered
using configuration interaction in a truncated Fock-space. The 2D
electron system is placed in a photon cavity with one mode of energy
$E_\mathrm{EM}$ and linear polarization in the $x$- or $y$-direction.
For the electron-photon interaction we retain both the para- and the dia-magnetic
terms without the rotating wave approximation, but consider the wavelength 
much larger than the size of the electron system 
\cite{Gudmundsson:2013.305,Gudmundsson12:1109.4728}.
The two parts of the electron-photon interaction are used since we consider the system 
both on and off resonance \cite{Jonasson2011:01}. We begin by using the
electron-photon coupling strength $g_\mathrm{EM}=0.05$ meV.
Not all types of cavities may admit an external perpendicular magnetic
field. We keep it in the model in order to take proper care of the spin
degree of freedom in the numerical calculations 
and to track possible effects of the coupling of the electron motion
along or perpendicular to the short quantum wire.

\begin{figure}[htbq]
      \includegraphics[width=0.44\textwidth,angle=0,bb=0 19 289 269,clip]{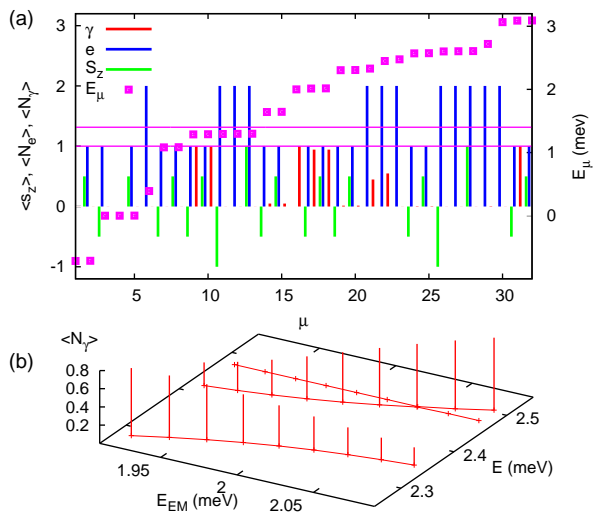}
      \caption{(Color online) (a) The lowest part of the energy spectrum $E_\mu$ (squares,
               units of meV) for the closed system vs.\ state number $\mu$. 
               The photon content $\langle N_\gamma\rangle$ of the states is indicated
               with vertical red bars, and the electron content $\langle N_e\rangle$ 
               and the spin
               $\langle s_z\rangle$ (units of $\hbar$)  with
               blue and green bars, respectively. 
               The photons are $y$-polarized with energy $E_\mathrm{EM}=2.0$ meV. 
               The two horizontal lines indicate the chemical potentials of 
               the biased leads that are coupled to the system to open it up
               to electrons, as discussed further in the text. 
               (b) The Rabi vacuum splitting of the two-electron states
               $|\breve{21})$ and $|\breve{22})$. The photon content is indicated with red
               bars. The two-electron state $|\breve{23})$ with vanishing photon content enters the 
               Rabi-splitting regime 
               and participates in the transport. $g_\mathrm{EM}=0.05$ meV. }
      \label{Fig-E}
\end{figure}

The energy spectrum of the closed system is displayed in Fig.\ \ref{Fig-E}(a)
together with information about the electron, photon, and spin content of the
lowest eigenstates for a photon field with $y$-polarization and energy 
$E_\mathrm{EM}$ chosen close to the confinement frequency $\hbar\Omega_0$.
We obtain a vacuum Rabi splitting for the two-electron state
containing one photon resulting in the Rabi pair ($|\breve{21})$,
$|\breve{22})$) seen in Fig. \ref{Fig-E}(b). We denote by
$|\breve{\mu})$ the composite many-body electron-photon eigenvectors
\cite{Gudmundsson:2013.305,Gudmundsson12:1109.4728}.

\textit{The closed system out of equilibrium}.--We now
consider a short classical electromagnetic pulse perturbing the closed system.
The time-evolution of the system is calculated by direct
integration of the Liouville-von Neumann equation for the density matrix
\cite{Gudmundsson03:161301,ANDP:ANDP201400048}.
\begin{figure}[htbq]
      \includegraphics[width=0.44\textwidth,angle=0,bb=0 0 184 234,clip]{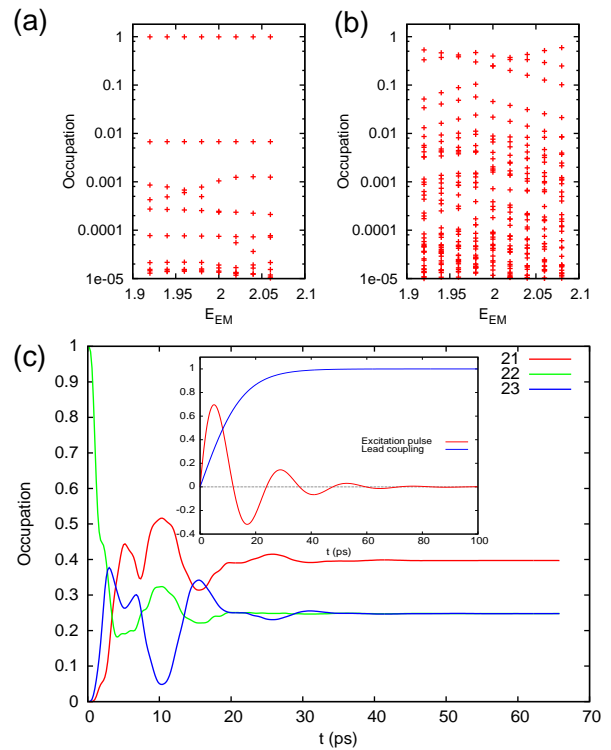}
      \caption{(Color online) For the closed system, (a) the occupation of states 
               $|\breve{\mu})$ for $x$-polarization
               and excitation in the $x$-direction. (b) The occupation for $y$-polarization
               and excitation in the $y$-direction. (c) The transient occupation for $y$-polarization
               and excitation in the $y$-direction for $E_\mathrm{EM}=2.0$ meV. 
               The inset shows the temporal part of the
               excitation pulse compared to the switching function for the lead coupling.
               The initial state is the lowest energy two-electron Rabi-split state with
               photon content $\gtrsim 0.5$, 
               for the $y$-polarization, but the two-electron ground state
               for the $x$-polarization.}
      \label{Fig-OPE}
\end{figure}
We start the time-evolution for the system in two different states,
with the cavity photons having either $x$- or $y$-polarization, 
and with the excitation pulse with the same polarization as the photons.
In the $x$-polarization case we use the two-electron ground state
$|\breve{6})$.  For the $y$-polarization we select the Rabi-split
state with the higher photon content ($\gtrsim 0.5$). 
After the excitation pulse has vanished the occupation
is constant and 
is seen in Fig.\ \ref{Fig-OPE}(a) and (b), for the two cases. 
The pulse is shown in the inset of Fig.\ \ref{Fig-OPE}(c) (red curve).

\begin{figure}[htbq]
      \includegraphics[width=0.44\textwidth,angle=0,bb=24 12 200 190,clip]{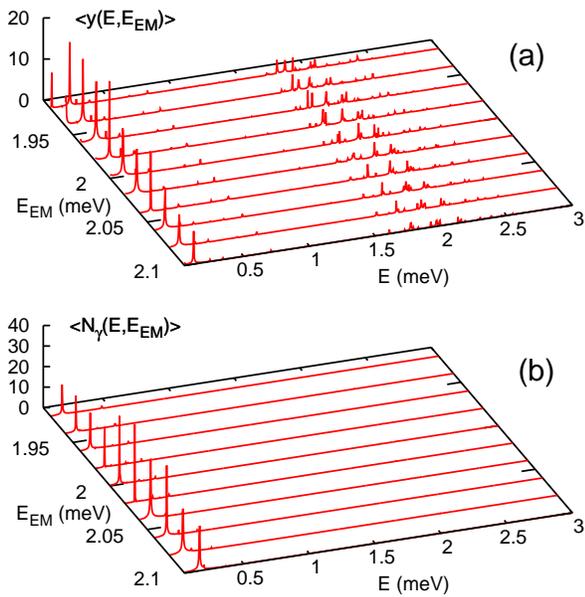}
      \caption{(Color online) The Fourier spectra in case of the closed system for
               (a) the center of mass coordinate $\langle y\rangle$, 
               and (b) the mean photon number $\langle N_\gamma\rangle$ for the initial 
               lowest energy Rabi-split two-electron state 
               with photon content $\gtrsim 0.5$. $g_\mathrm{EM}=0.05$ meV.}
      \label{Fig-Lok}
\end{figure}

The former excitation ($x$-polarization) gives
a gapped spectrum for which most transitions can be related to known
dipole active many-body states \cite{2040-8986-17-1-015201}.  
The latter excitation ($y$-polarization) is very close to a
resonance in the system and results in the activation of many transitions
visible in Fig.\ \ref{Fig-OPE}(b). More important is the fact that this type of low
frequency excitation pulse not only causes the occupation of the other Rabi-vacuum-split
state, i.\ e.\ $|\breve{21})$ together with  $|\breve{23})$, but also
a strong entanglement between the Rabi-vacuum-components (signaled
by large off-diagonal elements in the density matrix). The system
is far from an eigenstate and, as displayed in Fig.\ \ref{Fig-Lok}, it
shows very strong pure Rabi-oscillations in the mean photon number, Fig.\
\ref{Fig-Lok}(b), that are even present in the Fourier component of the
expectation value of the center of mass $y$ coordinate, see Fig.\ \ref{Fig-Lok}(a). 
If the photon energy is not in resonance with the confinement energy the
excitation spectrum of the mean values of the center of mass coordinates
is generally simpler for not too strong an excitation. This case was
already accounted for in Fig.\ \ref{Fig-OPE}(a), where an $x$-polarized
photon field is not in resonance with the electrons and higher states
above the ground state are only slightly occupied and no low energy
modes are excited.

\begin{figure}[htbq]
      \includegraphics[width=0.44\textwidth,angle=0,bb=0 6 198 196,clip]{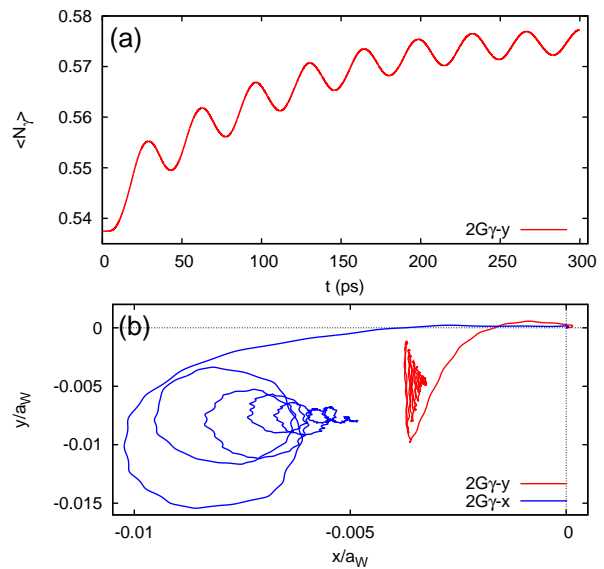}
      \caption{(Color online) For the open system (a) the mean photon number 
               $\langle N_\gamma\rangle$ for $y$-polarization, 
               and (b) the mean orbit of the center of mass for $x$-polarization (blue)
               and $y$-polarization (red) for the initial lowest energy Rabi-split 
               two-electron state with photon content $\gtrsim0.5$ ($y$-polarization),
               and the lowest energy two-electron one-photon state ($x$-polarization). 
               $E_\mathrm{EM}=2.0$ meV, $g_\mathrm{EM}=0.05$ meV.}
      \label{Fig-Opid}
\end{figure}

\textit{The open system}.--We have seen how an external electrical pulse
can be used to excite the system out of a many-body
eigenstate with a constant
photon number, into entangled states with an oscillating photon number. If we
increase the frequency of the excitation pulse the two Rabi-split states
get less entangled and smaller Rabi amplitude is observed. The question
is thus what happens if instead of applying an electrical pulse we open
up the system for transport of electrons through it. Can we expect to see
Rabi-oscillations then? To accomplish this we describe the coupling of the
system to two external parabolic semi-infinite leads with a non-Markovian
GME, selecting a time-dependent coupling function shown in the inset of
Fig.\ \ref{Fig-OPE}(c).  The coupling function has a similar timescale as
the external electrical pulse had.  The GME formalism with our spatially
dependent coupling of states in the leads and the system has been
described elsewhere \cite{Gudmundsson:2013.305,Gudmundsson12:1109.4728}
(here, the lead-system coupling strength is 0.5 meV and the lead temperature $T=0.5$ K).  
The chemical potentials 
of the left (L) and right (R) leads, $\mu_L=1.4$ meV and $\mu_R=1.1$ meV, respectively, 
are chosen to include 3 two-electron and 2 one-electron states
in the bias window, as is indicated in Fig.\ \ref{Fig-E}(a).  
Due to the geometry of the system the two-electron
states have low coupling to the leads, their charge densities being low in
the contact area of the central system.  In the case of a $y$-polarized
photon field approximately in resonance with the $y$-confinement potential
we observe small oscillations in the mean photon number seen in Fig.\
\ref{Fig-Opid}(a).  The oscillations are small since the GME-formalism
as applied here is only valid for weak contacts to the leads.
The frequency of the oscillations coincides with the Rabi-frequency observed in the 
closed system, and the Jaynes Cummings model \cite{Jaynes63:89} 
when the states $|\breve{6})$ and $|\breve{22})$ (see Fig.\ \ref{Fig-E}(a))
are taken as the ``atomic states'' with our electron-photon coupling strength, $g_\mathrm{EM}=0.05$ meV,
and photon energy $E_\mathrm{EM}=2.0$ meV. 

We are here describing different ways to 'excite' an electron-photon system confined by a
continuous potential. In order to describe correctly the strong electron-photon interaction
we need a large basis in the Fock space. As a byproduct we can see collective oscillations
emerging in the system opened up for transport, even in the weak coupling limit. 
In Fig.\ \ref{Fig-Opid}(b) we see the mean orbit of the center of mass of the two electrons for
the two linear polarizations of the photon field. In both cases the center of mass is shifted 
from the center of the system ($x=y=0$) to the left, as one of the electrons 
starts to seep slowly from the system into the right lead, performing revolutions
that are  synchronized with the oscillations of the photon number.
We see effects of the weak magnetic field, and the dissipation of energy to the leads.
The occupation of the initial two-electron state is getting less probable whereas 
lower energy one-electron states are gaining occupation probability.  
Fig.\ \ref{Fig-Opid}(b) shows how the system off resonance (blue curve) shows a simple spatial 
damped oscillation influenced by the magnetic field. In case of
the Rabi-resonance (red curve) the oscillation is almost entierly in the direction
dictated by the electrical component of the photon.  

\begin{figure}[htbq]
      \includegraphics[width=0.44\textwidth,angle=0,bb=50 59 388 221,clip]{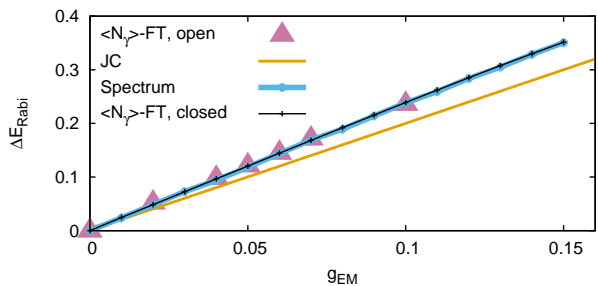}
      \caption{(Color online)  The energy of the Rabi-splitting as found from the 
               energy spectrum (Spectrum), the Fourier analysis of the oscillations in
               the mean photon number $\langle N_\gamma\rangle$ for the closed system
               ($\langle N_\gamma\rangle$-FT, closed), the open system 
               ($\langle N_\gamma\rangle$-FT, open), and the two-level Jaynes-Cummings model (JC). 
               $B=0.1$ T, $E_\mathrm{EM}=2.0$ meV.}
      \label{Rabi-g}
\end{figure}

The energy of the Rabi-splitting as a function of the coupling constant $g_\mathrm{EM}$ is 
compared for the open and the closed systems in Fig.\ \ref{Rabi-g}. 
The splitting for the open and the closed system agree within the accuracy of the 
numerical calculations. They are a bit higher than the value known for the 
two-level Jaynes-Cummings model $\Delta E_\mathrm{Rabi}^\mathrm{JC}=\sqrt{(\hbar\omega_r)^2+\delta^2}$,
with the detuning $\delta=7.44$ $\mu$eV and $\hbar\omega_r=2g_\mathrm{EM}$ for the vacuum
Rabi-oscillations. This can be expected for a multilevel model \cite{PhysRevA.52.2218}.

\begin{figure}[htbq]
      \includegraphics[width=0.44\textwidth,angle=0,bb=50 50 394 225,clip]{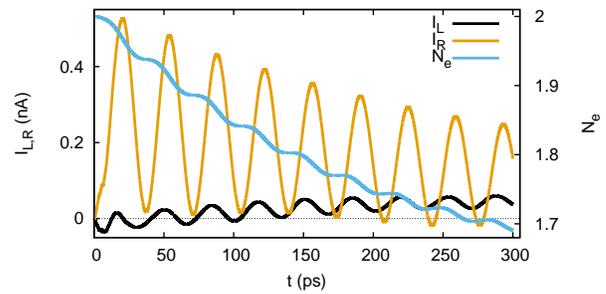}
      \caption{(Color online) For the open system the left and right currents (black and gold,
               respectively, left $y$-axis), and the
               mean number of electrons (blue, right $y$-axis) in case of 
               full entanglement for the Rabi-split states $|\breve{21})$ and $|\breve{22})$
               as initial states for a $y$-polarized photon field. $E_\mathrm{EM}=2.0$ meV.}
      \label{Fig-I-Rabi}
\end{figure}

Due to the restriction of the GME formalism to weak contacts the effects of the 
Rabi-oscillations on the current in the leads is minor. But if the system is
initially excited by an external electrical pulse before it is opened up for 
transport the initial state for the transport would be a highly entangled state
of the Rabi-split states and the current in the leads would reflect that, as can 
be seen in Fig.\ \ref{Fig-I-Rabi}.  
The oscillations in the current caused by the vacuum Rabi-oscillations decay with time
as the occupation of the two-electron Rabi-split pair of states get less probable as
charge enters and leaves the system.

\textit{Discussion and summary}.--Effort has been put into guaranteeing
the accuracy of the results presented here. The methods employed have
been based on a grid-free numerical approach in an appropriate
basis. We have used a so-called 'stepwise introduction of
model complexities with the necessary truncation' introduced elsewhere
\cite{Gudmundsson:2013.305}. The needed basis size was
in the range of 120 to 6000.

We show that even a weak contact of the central system to the external
leads causes collective oscillations of the electrons and the photons
in the system. Opposite to what happens in the closed system the collective oscillations 
in the open system can change their character as the state of the central system evolves
irreversibly in time. 
In order to describe the collective coupled oscillations of the strong interacting 
photons and electrons it is necessary to resort to large bases of electron states
and include both the para- and the dia-magnetic interactions. 

In the closed system we observe strong vacuum Rabi-oscillations in the photon content
when the photon frequency is close to the parabolic lateral confinement frequency  
and its polarization is in the perpendicular direction ($y$-direction). This situation favorites 
excitation by a low frequency perpendicular electrical pulse,
as the vacuum Rabi-splitting of the two-electron state is small. The excitation
pulse then effectively puts the system into an entangled state of the two Rabi-split states.    

The vacuum Rabi-oscillation is also seen in the open system under the same initial conditions
for the photon field, but its amplitude is small as the contacts to the leads are not very
effective in forming an entangled state of the Rabi-split states. This can be enhanced by
first exciting the system by an external electric pulse before it is opened up for electron
transport. 

There are two main reasons for selecting a parallel double quantum dot
system here.  First, they can capture states with two-electrons which have 
a rich spectrum of collective oscillations that can be excited by the transport. Second,
in order to observe Rabi- and collective oscillations we need the
system to be weakly contacted to the leads for the initial state to be
slowly decaying. For this purpose the two-electron states in parallel dots
are particularly convenient, as their coupling to the leads is highly
tunable \cite{Gudmundsson05:BT}.

%Acknowledgments
%
This work was financially supported by the Research Fund of the University of Iceland,
and the Icelandic Instruments Fund. We acknowledge also support from the computational 
facilities of the Nordic High Performance Computing (NHPC), and the Nordic network
NANOCONTROL, project No.: P-13053, and the Ministry of Science and Technology, Taiwan 
through Contract No. MOST 103-2112-M-239-001-MY3.

%
%---------------------------------------------
%
%
\bibliographystyle{apsrev4-1}
%

%
%
%----------------------------------------------------------------------------------------
%
\end{document}